\documentstyle[prd,aps,preprint,tighten,epsfig]{revtex}

\begin{document}

\title{Impacts of the observed $\theta^{}_{13}$ on the running behaviors
of Dirac and Majorana neutrino mixing angles and CP-violating phases}

\author{{\bf Shu Luo}
\thanks{E-mail: luoshu@xmu.edu.cn}}

\address{Department of Physics and Institute of Theoretical
Physics and Astrophysics, \\
Xiamen University, Xiamen, Fujian, 361005 China}

\author{{\bf Zhi-zhong Xing}
\thanks{E-mail: xingzz@ihep.ac.cn}}

\address{Institute of High Energy Physics, Chinese Academy of
Sciences, Beijing 100049, China \vspace{1cm}}

\maketitle

\begin{abstract}
The recent observation of the smallest neutrino mixing angle
$\theta^{}_{13}$ in the Daya Bay and RENO experiments motivates us
to examine whether $\theta^{}_{13} \simeq 9^\circ$ at the
electroweak scale can be generated from $\theta^{}_{13} =
0^\circ$ at a superhigh-energy scale via the radiative corrections. We
find that it is difficult but not impossible in the minimal
supersymmetric standard model (MSSM), and a relatively large
$\theta^{}_{13}$ may
have some nontrivial impacts on the running behaviors of the other
two mixing angles and CP-violating phases. In particular, we
demonstrate that the CP-violating phases play a crucial role in the
evolution of the mixing angles by using the one-loop
renormalization-group equations of the Dirac or Majorana neutrinos
in the MSSM. We also take the
``correlative" neutrino mixing pattern with $\theta^{}_{12} \simeq
35.3^\circ$, $\theta^{}_{23} = 45^\circ$ and $\theta^{}_{13} \simeq
9.7^\circ$ at a presumable flavor symmetry scale as an example to
illustrate that the three mixing angles can receive comparably small
radiative corrections and thus evolve to their best-fit values at
the electroweak scale if the CP-violating phases are properly
adjusted.
\end{abstract}


\newpage

\section{Introduction}

Since 1998, a number of successful neutrino oscillation experiments
have provided us with very compelling evidence that neutrinos are
massive and lepton flavors are mixed \cite{PDG}. The latest Daya Bay
\cite{Daya Bay} and RENO \cite{RENO} reactor antineutrino
oscillation experiments constitute another milestone, because their
results convince us that the smallest neutrino mixing angle
$\theta^{}_{13}$ is not really small: its best-fit value is about
$9^\circ$ \cite{Global Fit}. In comparison, the other two mixing
angles $\theta^{}_{12}$ and $\theta^{}_{23}$ are about $34^\circ$
and $45^\circ$, respectively \cite{Global Fit}. These three angles
appear in the standard parametrization of the $3\times 3$ lepton
flavor mixing matrix $V$,
\begin{equation}
V = \left( \matrix{ c^{}_{12}c^{}_{13} & s^{}_{12}c ^{}_{13} &
s^{}_{13} e^{-i\delta} \cr
-s^{}_{12}c^{}_{23}-c^{}_{12}s^{}_{23}s^{}_{13} e^{i\delta} &
c^{}_{12}c^{}_{23}-s^{}_{12}s^{}_{23}s^{}_{13} e^{i\delta} &
s^{}_{23}c^{}_{13} \cr s^{}_{12}s^{}_{23}
-c^{}_{12}c^{}_{23}s^{}_{13} e^{i\delta} & -c^{}_{12}s^{}_{23}
-s^{}_{12}c^{}_{23}s^{}_{13} e^{i\delta} & c^{}_{23}c^{}_{13}
} \right) \left ( \matrix{ e^{i \rho} & 0 & 0 \cr 0
& e^{i \sigma} & 0 \cr 0 & 0 & 1 \cr } \right ) \; ,
\end{equation}
where $c^{}_{ij} \equiv \cos\theta^{}_{ij}$ and $s^{}_{ij} \equiv
\sin\theta^{}_{ij}$ (for $ij=12, 23$ and $13$). The phase parameters
$\rho$ and $\sigma$ are usually referred to as the Majorana
CP-violating phases. If the massive neutrinos are the Dirac
particles, $\rho$ and $\sigma$ will have no physical meaning and can
be rotated away by rephasing the neutrino fields. At present the
three CP-violating phases remain unknown, but one of them (i.e., the
Dirac phase $\delta$) may hopefully be measured in the future
long-baseline neutrino oscillation experiments since
$\theta^{}_{13}$ is already known not to be very small. The strength
of leptonic CP violation in neutrino oscillations is governed by the
rephasing-invariant Jarlskog parameter ${\cal J} = c^{}_{12}
s^{}_{12} c^{}_{23} s^{}_{23} c^2_{13} s^{}_{13} \sin\delta$
\cite{J}, and that is why an appreciable value of $\theta^{}_{13}$
is a good news for us to explore leptonic CP violation in the near
future.

The fact that $\theta^{}_{13}$ is not as small as previously
expected motivates us to reconsider how it can be generated at the
tree level or by quantum corrections \cite{Xing2012}. In this paper
we shall follow a model-independent way to look at the impacts of a
relatively large $\theta^{}_{13}$ on the running behaviors of the
other two mixing angles and CP-violating phases for both Dirac and
Majorana neutrinos by using the one-loop renormalization-group
equations (RGEs) in the minimal supersymmetric standard model
(MSSM). Our purpose is to find out the conditions which should be
satisfied at a superhigh-energy scale where a flavor symmetry model
of neutrino masses can be built, in order to obtain a
phenomenologically favored neutrino mixing pattern at the
electroweak scale. In section II we reexamine the one-loop RGEs of
neutrino masses, mixing angles and CP-violating phase(s) by assuming
a nearly degenerate neutrino mass spectrum. The running behaviors of
the three mixing angles in the MSSM is numerically discussed in some
detail
\footnote{In this work we focus on the MSSM because the three
neutrino mixing angles can only receive much smaller radiative
corrections in the framework of the standard model
\cite{Majorana RGE,RGE,threshold effect}. Moreover, the
evolution of fermion masses and flavor mixing parameters in the
standard model may suffer from its vacuum stability problem if the
mass of the Higgs boson is about $125$ GeV \cite{XZZ2012}. }.
We pay particular attention to the crucial role of the CP-violating
phases in the RGE evolution. Section III is devoted to the analysis
of a special neutrino mixing pattern --- the so-called
``correlative" mixing pattern with $\theta^{}_{12} \simeq
35.3^\circ$, $\theta^{}_{23} = 45^\circ$ and $\theta^{}_{13} \simeq
9.7^\circ$ \cite{correlative mixing pattern}, which satisfy the sum
rule $\theta^{}_{12} + \theta^{}_{13} + \theta^{}_{23} = 90^\circ$,
as an example to illustrate that the three mixing angles can receive
comparably small radiative corrections and thus evolve to their
best-fit values at the electroweak scale if the CP-violating phases
are properly adjusted. A brief summary of our main results and
concluding remarks is given in section IV.

\section{One-loop RGEs for Dirac and Majorana Neutrinos}

\subsection{The Dirac case}

If the massive neutrinos are the Dirac particles, their Yukawa
coupling matrix $Y^{}_{\nu}$ must be extremely suppressed in
magnitude to reproduce the light neutrino masses of ${\cal O}(1)$ eV
or smaller at low energy scales. In the MSSM, the running of
$Y^{}_{\nu}$ from the electroweak scale $\Lambda^{}_{\rm EW}$ to a
superhigh-energy scale $\Lambda$ is governed by the one-loop RGE
\cite{Dirac RGE,Dirac RGE tau}
\begin{equation}
16\pi^2 \frac{{\rm d}\omega}{{\rm d}t} \; = \; 2 \alpha^{}_{\rm D}
\omega + \left [ \left ( Y^{}_{l} Y^{\dagger}_{l} \right ) \omega +
\omega \left ( Y^{}_{l} Y^{\dagger}_{l} \right ) \right ] \; ,
\end{equation}
where $\omega \equiv Y^{}_{\nu} Y^{\dagger}_{\nu}$,
$t \equiv \ln \left ( \mu / \Lambda \right )$ with $\mu$
being an arbitrary renormalization scale between $\Lambda^{}_{\rm
EW}$ and $\Lambda$, $Y^{}_{l}$ is the charged-lepton Yukawa coupling
matrix, and $\alpha^{}_{\rm D} \approx - 0.6 g^{2}_{1} - 3 g^{2}_{2} + 3
y^{2}_{t}$. Here $g^{}_{1}$ and $g^{}_{2}$ are the gauge couplings,
$y^{}_{t}$ stands for the top-quark Yukawa coupling. In writing out
Eq. (2), those tiny terms of ${\cal O}(\omega^{2}_{})$ have been safely
omitted.

One may use Eq. (2) to derive the explicit RGEs of the three
neutrino masses and four mixing parameters. The results can be found
either in Ref. \cite{Dirac RGE} where the standard parametrization
of $V$ is adopted, or in Ref. \cite{Dirac RGE tau} where the more
convenient Fritzsch-Xing parametrization of $V$ \cite{tau-dominant}
is used. To see the appreciable running effects, here we assume that
the masses of the three neutrinos are nearly degenerate. Since
neglecting the small contributions of $y^{}_e$ and $y^{}_\mu$ in Eq.
(2) is always a good approximation, we shall also do so in our
calculations. Given the near mass degeneracy $m^{}_{1} \approx
m^{}_{2} \approx m^{}_{3}$ together with the standard
parametrization of $V$ in Eq. (1), the RGEs of $m^{}_i$ (for
$i=1,2,3$) turn out to be
\begin{eqnarray}
\frac{{\rm d} m^{}_1}{{\rm d} t} & \approx & \frac{m^{}_1}{16\pi^2}
\left [ \alpha^{}_{\rm D}  + y^{2}_{\tau} \left (s^{2}_{12}
s^{2}_{23} - 2 c^{}_{\delta} c^{}_{12} c^{}_{23} s^{}_{12} s^{}_{23}
s^{}_{13} + {\cal O} (s^{2}_{13}) \right ) \right ] \; ,
\nonumber \\
\frac{{\rm d} m^{}_2}{{\rm d} t} & \approx & \frac{m^{}_2}{16\pi^2}
\left [ \alpha^{}_{\rm D} + y^{2}_{\tau} \left ( c^{2}_{12}
s^{2}_{23} + 2 c^{}_{\delta } c^{}_{12} c^{}_{23} s^{}_{12}
s^{}_{23} s^{}_{13}  + {\cal O} (s^{2}_{13}) \right ) \right ] \; ,
\nonumber \\
\frac{{\rm d} m^{}_3}{{\rm d} t} & \approx & \frac{m^{}_3}{16\pi^2}
\left [ \alpha^{}_{\rm D}  + y^{2}_{\tau} c^{2}_{23}  + {\cal O}
(s^{2}_{13}) \right ] \; ,
\end{eqnarray}
where $c^{}_{\delta} \equiv \cos\delta$ and $y^2_\tau = m^2_\tau
(1+\tan^2\beta)/v^2 \simeq (1+\tan^2\beta) \times 10^{-4}$ with
$v \simeq 174$ GeV and
$\tan\beta$ being the ratio of the vacuum expectation values of the
two Higgs doublets in the MSSM. The RGEs of $\theta^{}_{ij}$ (for
$ij =12, 23, 13$) are found to be
\begin{eqnarray}
\frac{{\rm d} \theta^{}_{12}}{{\rm d} t} & \approx & -
\frac{y^{2}_{\tau}}{8 \pi^2} \left [ \frac{m^{2}_{1}}{\Delta
m^{2}_{21}} \left ( c^{}_{12} s^{}_{12} s^{2}_{23} -
\cos2\theta^{}_{12} c^{}_{23} s^{}_{23} s^{}_{13} c^{}_{\delta}
\right ) - \frac{m^{2}_{1}}{\Delta m^{2}_{32}} c^{}_{23} s^{}_{23}
s^{}_{13} c^{}_{\delta}  + {\cal O} (s^{2}_{13}) \right ] \; ,
\nonumber \\
\frac{{\rm d} \theta^{}_{23}}{{\rm d} t} & \approx & -
\frac{y^{2}_{\tau}}{8 \pi^2} \frac{m^{2}_{1}}{\Delta m^{2}_{32}}
c^{}_{23} s^{}_{23}  + {\cal O} (s^{2}_{13}) \; ,
\nonumber \\
\frac{{\rm d} \theta^{}_{13}}{{\rm d} t} & \approx & -
\frac{y^{2}_{\tau}}{8 \pi^2} \frac{m^{2}_{1}}{\Delta m^{2}_{32}}
c^{2}_{23} c^{}_{13} s^{}_{13}  + {\cal O} (s^{2}_{13}) \; .
\end{eqnarray}
In addition, the RGE of the CP-violating phase $\delta$ can be written as
\begin{eqnarray}
\frac{{\rm d} \delta}{{\rm d} t} & \approx & - \frac{y^{2}_{\tau}}{8
\pi^2} \left ( \frac{m^{2}_{1}}{\Delta m^{2}_{21}} +
\frac{m^{2}_{1}}{\Delta m^{2}_{32}} \cos2\theta^{}_{12} \right )
\frac{c^{}_{23} s^{}_{23} s^{}_{13} s^{}_{\delta}}{c^{}_{12}
s^{}_{12}}  + {\cal O} (s^{2}_{13}) \; ,
\end{eqnarray}
where $s^{}_{\delta} \equiv \sin\delta$ is defined. Note that the
full RGE of $\delta$ actually contains the terms which are inversely
proportional to $s^{}_{13}$, but they are negligible in Eq. (5) for
two reasons: (a) they are not significantly enhanced just because
$\theta^{}_{13}$ is not very small, as observed in the recent Daya
Bay and RENO experiments; and (b) they are significantly suppressed
by the factor $\Delta m^{2}_{21} / \Delta m^{2}_{32}$ in the case of
$m^{}_{1} \approx m^{}_{2} \approx m^{}_{3}$ under discussion.
Finally, the RGE of the Jarlskog invariant ${\cal J}$ is
\begin{eqnarray}
\frac{\rm d {\cal J}}{{\rm d}t} & \approx & -
\frac{y^2_\tau}{8\pi^2} {\cal J} \left \{ \frac{m^{2}_{1}}{\Delta
m^{2}_{21}} \left [ \cos2\theta^{}_{12} s^{2}_{23} + 4 c^{}_{12}
s^{}_{12} c^{}_{23} s^{}_{23} s^{}_{13} c^{}_{\delta} \right ] +
\frac{m^{2}_{1}}{\Delta m^{2}_{32}} \left ( 3 c^{2}_{23} - 1 \right
) + {\cal O} (s^{2}_{13}) \right \} \; .
\end{eqnarray}
Some discussions are in order.
\begin{itemize}
\item      Eq. (4) clearly tells us that the one-loop RGE of
$\theta^{}_{12}$ is dominated by the leading term $\displaystyle -
\frac{y^{2}_{\tau}}{8 \pi^2} \frac{m^{2}_{1}}{\Delta m^{2}_{21}}
c^{}_{12} s^{}_{12} s^{2}_{23} $. As a consequence of $\Delta
m^{2}_{21} \ll |\Delta m^{2}_{32}|$, the solar mixing angle
$\theta^{}_{12}$ is more sensitive to radiative corrections than the
other two angles $\theta^{}_{13}$ and $\theta^{}_{23}$. Our
numerical analysis shows that $\theta^{}_{12}$ may undergo an
increase of about $15^{\circ}$ to $20^{\circ}_{}$ in the MSSM with
$\tan\beta =10$ (denoted as ``MSSM10" hereafter for short) or an
increase of about $20^{\circ}$ to $25^{\circ}_{}$ in the MSSM with
$\tan\beta =50$ (denoted as ``MSSM50" hereafter for short) for
arbitrary values of $\delta$, if it evolves from a flavor symmetry
scale $\Lambda^{}_{\rm FS} \sim 10^{14}_{}$ GeV down to the
electroweak scale $\Lambda^{}_{\rm EW} \sim 10^{2}$ GeV.

\item      Running from $\Lambda^{}_{\rm FS}$ down to
$\Lambda^{}_{\rm EW}$, the values of $\theta^{}_{23}$ and
$\theta^{}_{13}$ can eithe increase or decrease, depending on the
sign of $\Delta m^{2}_{32}$. Given the MSSM10 case for example,
$\theta^{}_{23}$ changes about $1.5^{\circ}_{}$ while
$\theta^{}_{13}$ changes about $0.2^{\circ}_{}$ from
$\Lambda^{}_{\rm FS}$ to $\Lambda^{}_{\rm EW}$ (or vice versa). In
the MSSM50 case, $\theta^{}_{23}$ may change about $12^{\circ}_{}$
while $\theta^{}_{13}$ changes about $6^{\circ}$ from
$\Lambda^{}_{\rm FS}$ to $\Lambda^{}_{\rm EW}$ (or vice versa). Note
that the radiative corrections to $\theta^{}_{23}$ and
$\theta^{}_{13}$ are almost independent of the CP-violating phase
$\delta$, as one can easily see from Eq. (4).

\item      The one-loop RGE of ${\cal J}$ is proportional to ${\cal J}$
itself, and that of $\delta$ is proportional to $\sin\delta$. Hence
the evolution of $\cal J$ or $\delta$ does not undergo a sign flip.
In other words, CP violation is an intrinsic property of the lepton
flavor structure: if it is present (or absent) at a given energy
scale, it must be present (or absent) at any other energy scales.
Evolving from $\Lambda^{}_{\rm FS}$ down to $\Lambda^{}_{\rm EW}$,
$|{\cal J}|$ and $|\delta|$ (for $- \pi \leq \delta \leq \pi$) will
increase in the MSSM.
\end{itemize}
In the next subsection we shall see that a relatively large
$\theta^{}_{13}$ has much more interesting phenomenological
consequences provided the massive neutrinos are the Majorana
particles.

\subsection{The Majorana case}

The masses of the Majorana neutrinos are believed to be attributed
to some underlying new physics at a superhigh-energy scale $\Lambda$
(e.g., via the canonical seesaw mechanism \cite{SS}). But this kind of new
physics can all point to the unique dimension-5 Weinberg operator
for the neutrino masses in an effective field theory after the
corresponding heavy degrees of freedom are integrated out
\cite{Weinberg1979}. In the MSSM, such a dimension-5 operator reads
\begin{eqnarray}
\frac{{\cal L}^{}_{\rm d=5}}{\Lambda} & = & \frac{1}{2} \;
\overline{\ell^{}_{\rm L}} H^{}_{2} \cdot \kappa \cdot H^{T}_{2}
\ell^{c}_{\rm L} \; + \; {\rm h.c.} \; ,
\end{eqnarray}
where $\Lambda$ denotes the cutoff scale, $\ell^{}_{\rm L}$ stands
for the left-handed lepton doublet, $H^{}_2$ is one of the MSSM
Higgs doublets, and $\kappa$ represents the effective neutrino
coupling matrix. One may obtain the effective Majorana neutrino mass
matrix $M^{}_{\nu} = \kappa v^2 \tan^2\beta/(1 + \tan^2\beta)$ after
spontaneous gauge symmetry breaking. The cutoff scale $\Lambda$
actually stands for the scale of new physics, such as the mass scale
of the heavy Majorana neutrinos in the canonical seesaw mechanism
\cite{SS}. The evolution of $\kappa$ from $\Lambda$ down to the
electroweak scale $\Lambda^{}_{\rm EW}$ is formally independent of
any details of the relevent model from which $\kappa$ is derived.
Below $\Lambda$ the scale dependence of $\kappa$ is described by
\begin{equation}
16\pi^2 \frac{{\rm d}\kappa}{{\rm d}t} \; = \; \alpha^{}_{\rm M}
\kappa + \left [ \left ( Y^{}_{l} Y^{\dagger}_{l} \right ) \kappa +
\kappa \left ( Y^{}_{l} Y^{\dagger}_{l} \right )^{T}_{} \right ] \;
\end{equation}
at the one-loop level in the MSSM \cite{Majorana RGE}, where
$\alpha^{}_{\rm M} \approx - 1.2 g^{2}_{1} - 6 g^{2}_{2} + 6
y^{2}_{t}$.

One may use Eq. (8) to derive the explicit RGEs of the three
neutrino masses and six flavor mixing parameters \cite{RGE,threshold
effect}. Given an approximate mass degeneracy of the three neutrinos
together with the standard parametrization of $V$ in Eq. (1), the
RGEs of $m^{}_i$ (for $i=1,2,3$) turn out to be
\begin{eqnarray}
\frac{{\rm d} m^{}_1}{{\rm d} t} & \approx & \frac{m^{}_1}{16\pi^2}
\left [ \alpha^{}_{\rm M}  + 2 y^{2}_{\tau} \left (s^{2}_{12}
s^{2}_{23} - 2 c^{}_{\delta} c^{}_{12} c^{}_{23} s^{}_{12} s^{}_{23}
s^{}_{13}  + {\cal O} (s^{2}_{13}) \right ) \right ] \; ,
\nonumber \\
\frac{{\rm d} m^{}_2}{{\rm d} t} & \approx & \frac{m^{}_2}{16\pi^2}
\left [ \alpha^{}_{\rm M} + 2 y^{2}_{\tau} \left ( c^{2}_{12}
s^{2}_{23} + 2 c^{}_{\delta } c^{}_{12} c^{}_{23} s^{}_{12}
s^{}_{23} s^{}_{13}  + {\cal O} (s^{2}_{13}) \right ) \right ] \; ,
\nonumber \\
\frac{{\rm d} m^{}_3}{{\rm d} t} & \approx & \frac{m^{}_3}{16\pi^2}
\left [ \alpha^{}_{\rm M}  + 2 y^{2}_{\tau} c^{2}_{23} + {\cal O}
(s^{2}_{13}) \right ] \; .
\end{eqnarray}
The RGEs of $\theta^{}_{ij}$ (for $ij =12, 23, 13$) are found to be
\begin{eqnarray}
\frac{{\rm d} \theta^{}_{12}}{{\rm d} t} & \approx & -
\frac{y^{2}_{\tau}}{4 \pi^2} \left \{ \frac{m^{2}_{1}}{\Delta
m^{2}_{21}} s^{}_{23} \left [ \left ( c^{}_{12} s^{}_{12} s^{}_{23} -
\cos2\theta^{}_{12} c^{}_{23} s^{}_{13} c^{}_{\delta}
\right ) c^{}_{(\rho - \sigma)} + c^{}_{23} s^{}_{13}
s^{}_{\delta} s^{}_{(\rho - \sigma)} \right ] c^{}_{(\rho - \sigma)}
\right.
\nonumber\\
& & ~~~ \left. - \frac{m^{2}_{1}}{\Delta m^{2}_{32}} c^{}_{23}
s^{}_{23} s^{}_{13} \left ( s^{2}_{12} c^{}_{(\delta + \rho)}
c^{}_{\rho} + c^{2}_{12} c^{}_{(\delta + \sigma)} c^{}_{\sigma}
\right )  + {\cal O} (s^{2}_{13}) \right \} \; ,
\nonumber \\
\nonumber \\
\frac{{\rm d} \theta^{}_{23}}{{\rm d} t} & \approx & -
\frac{y^{2}_{\tau}}{4 \pi^2} \frac{m^{2}_{1}}{\Delta m^{2}_{32}}
c^{}_{23} \left [s^{}_{23} \left ( s^{2}_{12} c^{2}_{\rho}
+c^{2}_{12} c^{2}_{\sigma} \right ) - \frac{1}{2} c^{}_{12}
s^{}_{12} c^{}_{23} s^{}_{13} \left ( c^{}_{(\delta + 2 \rho)} -
c^{}_{(\delta + 2 \sigma)} \right ) + {\cal O} (s^{2}_{13})\right ]
\; ,
\nonumber \\
\nonumber \\
\frac{{\rm d} \theta^{}_{13}}{{\rm d} t} & \approx &
\frac{y^{2}_{\tau}}{8 \pi^2} \frac{m^{2}_{1}}{\Delta m^{2}_{32}}
c^{}_{23} c^{}_{13} \left [ c^{}_{12} s^{}_{12} s^{}_{23} \left (
c^{}_{(\delta + 2 \rho)} - c^{}_{(\delta + 2 \sigma)} \right ) \right.
\nonumber\\[2mm]
& & \left. ~~~ - 2
c^{}_{23} s^{}_{13} \left ( c^{2}_{12} c^{2}_{(\delta + \rho)} +
s^{2}_{12} c^{2}_{(\delta + \sigma)} \right ) + {\cal O}
(s^{2}_{13}) \right ] \; ,
\end{eqnarray}
in which $c^{}_{x} \equiv \cos x$ and $s^{}_{x} \equiv \sin x$ (for
$x = \delta, \; \rho, \; \sigma, \; \rho-\sigma, \; \delta+\rho, \;
\delta+\sigma, \; \delta+2\rho, \; \delta+2\sigma$). The RGE of the
three CP-violating phases $\delta$, $\rho$ and $\sigma$ can be
written as
\begin{eqnarray}
\frac{{\rm d} \delta}{{\rm d} t} & \approx & - \frac{y^{2}_{\tau}}{4
\pi^2} \left \{ \frac{m^{2}_{1}}{\Delta m^{2}_{21}} s^{}_{23} \left
[ \left ( s^{}_{23} - \frac{\cos2\theta^{}_{12} c^{}_{23} s^{}_{13}
c^{}_{\delta}}{c^{}_{12} s^{}_{12}} \right ) c^{}_{(\rho - \sigma)}
+ \frac{c^{}_{23} s^{}_{13} s^{}_{\delta}}{c^{}_{12} s^{}_{12}}
s^{}_{(\rho - \sigma)} + {\cal O} (s^{2}_{13}) \right ] s^{}_{(\rho
- \sigma)} \right.
\nonumber\\
& & ~~~~~~~ \left. - \frac{m^{2}_{1}}{\Delta m^{2}_{32}}
s^{-1}_{13} \left [ \frac{1}{2} c^{}_{12} s^{}_{12} c^{}_{23}
s^{}_{23} \left ( s^{}_{(\delta + 2 \rho)} - s^{}_{(\delta +
2 \sigma)} \right ) + \left ( c^{}_{\rho} s^{}_{\rho} c^{2}_{12}
+ c^{}_{\sigma} s^{}_{\sigma} s^{2}_{12} \right ) c^{2}_{23} s^{}_{13}
\right. \right. \nonumber\\[2mm]
& & ~~~~~~~~~~~~~~~~~~~~~ \left. \left. + \left ( c^{}_{(\delta -
\rho)} s^{}_{(\delta - \rho)} s^{2}_{12} +c^{}_{(\delta - \sigma)}
s^{}_{(\delta - \sigma)} c^{2}_{12} \right ) \cos2\theta^{}_{23}
s^{}_{13} + {\cal O} (s^{2}_{13}) \right ] \right \} \; ,
\nonumber \\
\nonumber \\
\frac{{\rm d} \rho}{{\rm d} t} & \approx & \frac{y^{2}_{\tau}}{4
\pi^2} \left \{ \frac{m^{2}_{1}}{\Delta m^{2}_{21}} s^{}_{23}
c^{2}_{12} \left [ \left ( s^{}_{23} - \frac{\cos2\theta^{}_{12}
c^{}_{23} s^{}_{13} c^{}_{\delta}}{c^{}_{12} s^{}_{12}} \right )
c^{}_{(\rho - \sigma)} + \frac{c^{}_{23} s^{}_{13}
s^{}_{\delta}}{c^{}_{12} s^{}_{12}} s^{}_{(\rho - \sigma)} + {\cal
O} (s^{2}_{13}) \right ] s^{}_{(\rho - \sigma)} \right.
\nonumber\\
& & ~~~~~~ \left. - \frac{m^{2}_{1}}{\Delta m^{2}_{32}} s^{-1}_{13}
\left [ \left ( c^{}_{(\delta - \rho)} s^{}_{(\delta - \rho)}
s^{2}_{12} +c^{}_{(\delta - \sigma)} s^{}_{(\delta - \sigma)}
c^{2}_{12} \right ) \cos2\theta^{}_{23} s^{}_{13} + {\cal O}
(s^{2}_{13}) \right ] \right \} \; ,
\nonumber \\
\nonumber \\
\frac{{\rm d} \sigma}{{\rm d} t} & \approx & \frac{y^{2}_{\tau}}{4
\pi^2} \left \{ \frac{m^{2}_{1}}{\Delta m^{2}_{21}} s^{}_{23}
s^{2}_{12} \left [ \left (s^{}_{23} - \frac{\cos2\theta^{}_{12}
c^{}_{23} s^{}_{13} c^{}_{\delta}}{c^{}_{12} s^{}_{12}} \right )
c^{}_{(\rho - \sigma)} + \frac{c^{}_{23} s^{}_{13}
s^{}_{\delta}}{c^{}_{12} s^{}_{12}} s^{}_{(\rho - \sigma)} + {\cal
O} (s^{2}_{13}) \right ] s^{}_{(\rho - \sigma)} \right.
\nonumber\\
& & ~~~~~~ \left. - \frac{m^{2}_{1}}{\Delta m^{2}_{32}} s^{-1}_{13}
\left [ \left ( c^{}_{(\delta - \rho)} s^{}_{(\delta - \rho)}
s^{2}_{12} +c^{}_{(\delta - \sigma)} s^{}_{(\delta - \sigma)}
c^{2}_{12} \right ) \cos2\theta^{}_{23} s^{}_{13} + {\cal O}
(s^{2}_{13}) \right ] \right \} \; .
\end{eqnarray}
In addition, the RGE of ${\cal J}$ is obtained as follows:
\begin{eqnarray}
\frac{\rm d}{{\rm d}t} \; {\cal J} & \approx & -
\frac{y^2_\tau}{8\pi^2} \left \{ \frac{m^{2}_{1}}{\Delta m^{2}_{21}} \left
[ {\cal J} \cos2\theta^{}_{12} s^{2}_{23} -
\cos^{2}_{}2\theta^{}_{12} c^{2}_{23} s^{2}_{23} c^{2}_{13}
s^{2}_{13} c^{}_{\delta} s^{}_{\delta} \right ] \right. \nonumber\\
& & ~~~~~~ \left. + \; \frac{m^{2}_{1}}{\Delta m^{2}_{32}} {\cal
J} \cos2\theta^{}_{23} + {\cal O} (s^{3}_{13}) \right \} \; .
\end{eqnarray}
Because of ${\cal J} \propto \sin\delta$, the running of $\cal J$ is
also proportional to $\sin\delta$. This situation is similar to the
evolution of $\cal J$ in the Dirac case. But now the evolution of
$\delta$ is nonlinearly entangled with the evolution of $\rho$ and
$\sigma$ as shown in Eq. (11), so the Majorana case is more
complicated than the Dirac case.

The running behaviors of the three mixing angles and three
CP-violating phases can be very different for a very small
$\theta^{}_{13}$ and for a relative large $\theta^{}_{13}$. In
particular, the CP-violating phases play a crucial role in the RGEs.
Let us elaborate on this point in the following.

\subsubsection{The running behaviors of the three mixing angles}

Eq. (10) shows that the RGE running behaviors of the three neutrino
mixing angles are strongly dependent on the three CP-violating
phases. As for the Majorana neutrinos, the radiative corrections to
the three mixing angles can be adjusted by choosing different values
of the CP-violating phases $\delta$, $\rho$ and $\sigma$. To
illustrate this observation, let us carry out an explicit numerical
analysis. We choose $\theta^{}_{12} = 34^\circ$, $\theta^{}_{23} =
46^\circ$, $\theta^{}_{13} = 9^\circ$, $\delta = 90^\circ$ and
$\sigma = 30^\circ$ as the typical inputs at $\Lambda^{}_{\rm EW}$
and allow $\rho$ to vary from $0^\circ$ to $180^\circ$. Then we look
at their numerical evolution to $\Lambda^{}_{\rm FS}$ via the RGEs.
FIGs. 1 and 2 show the possible ranges of the three mixing angles at
$\mu > \Lambda^{}_{\rm EW}$ (gray areas) for both normal and
inverted neutrino mass hierarchies, where $m^{}_{1} \sim 0.2$ eV has
typically been input at $\Lambda^{}_{\rm EW}$. The dashed ($\Delta
m^{2}_{23} >0$) and dotted-dashed ($\Delta m^{2}_{23} <0$) lines in
these figures represent the corresponding running behaviors of the
three mixing angles for the Dirac neutrinos (with the same input
values of $\theta^{}_{12}$, $\theta^{}_{23}$, $\theta^{}_{13}$ and
$\delta$).

The fact that the RGE of $\theta^{}_{12}$ is dominated by
the term $\displaystyle - \frac{y^{2}_{\tau}}{8 \pi^2}
\frac{m^{2}_{1}}{\Delta m^{2}_{21}} c^{}_{12} s^{}_{12} s^{2}_{23}
c^{2}_{(\rho - \sigma)} $ implies that the magnitude of the radiative
correction to $\theta^{}_{12}$ depends strongly on the phase
difference $(\rho - \sigma)$. Hence $\theta^{}_{12}$ is most sensitive
to the RGE effect when $\rho \simeq \sigma$ holds. FIG. 1 shows that
in the MSSM10 case the resulting $\theta^{}_{12}$ at
$\Lambda^{}_{\rm FS}$ lies in a wide range (from $10^\circ$ to
$35^\circ$ associated with the variation of $\rho$). While in the
MSSM50 case the resulting $\theta^{}_{12}$ at $\Lambda^{}_{\rm FS}$
lies in a wider range (from $7^\circ$ to $55^\circ$ if $\Delta
m^{2}_{23} >0$, or from $2^\circ$ to $31^\circ$ if $\Delta
m^{2}_{23} <0$). If all the three CP-violating phases are freely
adjusted, the allowed range of $\theta^{}_{12}$ at $\Lambda^{}_{\rm
FS}$ will become much wider (from $0.5^\circ$ to $62^\circ$ if
$\Delta m^{2}_{23} >0$, or from $2^\circ$ to $45^\circ$ if $\Delta
m^{2}_{23} <0$).

Running from $\Lambda^{}_{\rm FS} \sim 10^{14}_{}$ GeV down to
$\Lambda^{}_{\rm EW}$, the mixing angles $\theta^{}_{23}$ and
$\theta^{}_{13}$ receive less significant radiative corrections.
They may change one or two degrees in the MSSM10 case, as shown in
FIG. 1. In the MSSM50 case shown in FIG. 2, $\theta^{}_{23}$ may
increase or decrease in the range of $22^{\circ}$ to $35^{\circ}$ between
the scales $\Lambda^{}_{\rm EW}$ and $\Lambda^{}_{\rm FS}$, whereas
the change of $\theta^{}_{13}$ lies in the range of
$4.1^\circ$ to $22.5^\circ$ (for $\Delta m^{2}_{23} >0$)
or in the range of $13.0^\circ$ to $30.3^\circ$ (for $\Delta
m^{2}_{23} <0$). If all the three CP-violating phases vary freely,
the resulting $\theta^{}_{23}$ at $\Lambda^{}_{\rm FS}$
lies in the range of $7.5^\circ$ to $45.5^\circ$
(for $\Delta m^{2}_{23} >0$) or in the range of $46.5^\circ$ to
$89^\circ$ (for $\Delta m^{2}_{23} <0$), while
the resulting $\theta^{}_{13}$ at $\Lambda^{}_{\rm FS}$ lies in
the range of $2^\circ$ to $23^\circ$ (for $\Delta m^{2}_{23} >0$)
or in the range of $7.5^\circ$ to $82^\circ$ (for $\Delta m^{2}_{23} <0$).
One can see that it is impossible
to generate $\theta^{}_{13} \simeq 9^\circ$ at $\Lambda^{}_{\rm
EW}$ from $\theta^{}_{13} \simeq 0^\circ$ at
$\Lambda^{}_{\rm FS}$ via the radiative corrections. This
observation is true even in the MSSM50 case.

We have seen that the values of the three CP-violating phases are
crucial for the evolution of the three mixing angles. A very special
case is $(\rho - \sigma) \simeq \pm 90^\circ$, which leads us to
\begin{eqnarray}
\frac{{\rm d} \theta^{}_{12}}{{\rm d} t} & \approx &
\frac{y^{2}_{\tau}}{4 \pi^2}  \frac{m^{2}_{1}}{\Delta m^{2}_{32}}
c^{}_{23} s^{}_{23} s^{}_{13} \left ( s^{2}_{12} c^{}_{(\delta +
\rho)} c^{}_{\rho} + c^{2}_{12} s^{}_{(\delta + \rho)} s^{}_{\rho}
\right ) \; ,
\nonumber \\
\nonumber \\
\frac{{\rm d} \theta^{}_{23}}{{\rm d} t} & \approx & -
\frac{y^{2}_{\tau}}{4 \pi^2} \frac{m^{2}_{1}}{\Delta m^{2}_{32}}
c^{}_{23} \left [ s^{}_{23} \left ( s^{2}_{12} c^{2}_{\rho}
+c^{2}_{12} s^{2}_{\rho} \right ) -  c^{}_{12} s^{}_{12} c^{}_{23}
s^{}_{13} c^{}_{(\delta + 2 \rho)} \right ] \; ,
\nonumber \\
\nonumber \\
\frac{{\rm d} \theta^{}_{13}}{{\rm d} t} & \approx &
\frac{y^{2}_{\tau}}{4 \pi^2} \frac{m^{2}_{1}}{\Delta m^{2}_{32}}
c^{}_{23} c^{}_{13} \left [ c^{}_{12} s^{}_{12} s^{}_{23}
c^{}_{(\delta + 2 \rho)} - c^{}_{23} s^{}_{13} \left ( c^{2}_{12}
c^{2}_{(\delta + \rho)} + s^{2}_{12} s^{2}_{(\delta + \rho)} \right
) \right ] \; .
\end{eqnarray}
Note that the term proportional to $m^{2}_{1} / \Delta m^{2}_{21}$
in the RGE of $\theta^{}_{12}$ in Eq. (10) is suppressed
by $\cos(\rho -\sigma) \simeq 0$ in this special case,
and thus it has been omitted from Eq. (13). The three mixing angles
may therefore receive comparably small radiative corrections for a modest
value of $\tan\beta$ (e.g., in the MSSM10 case).
This observation was not noticed in the literature simply because
$\theta^{}_{13}$ used to be assumed to be very small
\cite{RGE,threshold effect}. If $\tan\beta$ is sufficiently large
(e.g., in the MSSM50 case), however, the phase difference
$(\rho - \sigma)$ will be able to quickly run away from its initial value
$(\rho -\sigma) \sim \pm 90^\circ$ due to the significant radiative
corrections, implying that Eq. (13) is no more a good approximation
of Eq. (10).

\subsubsection{The radiative generation of the CP-violating phases}

It is well known that one CP-violating phase can be generated from
another \cite{radiative generation}, simply because they are entangled
in the RGEs. An especially interesting example is the
Dirac phase $\delta$, which measures the strength of CP violation
in neutrino oscillations at the electroweak scale, can be radiatively
generated from the nonzero Majorana phases $\rho$ and $\sigma$ at
a superhigh-energy scale. To illustrate, we present two numerical
examples in Tables I and II to show that it is possible to
radiatively generate $\delta$ and one of the two Majorana phases
from the other Majorana phase. If $\theta^{}_{13}$ is very small,
however, the running of $\delta$ can be significantly enhanced by
the terms that are inversely proportional to $\sin\theta^{}_{13}$.
In the MSSM10 case it has been found that even $\delta = 90^\circ$ can be
radiatively generated if $\theta^{}_{13} \simeq 1^\circ$ is taken
\cite{radiative generation}. In our numerical calculation we
require $\theta^{}_{13} \simeq 9^\circ$ at $\Lambda_{\rm EW}$.
Thanks to the RGE running effects, we find that
$-30^\circ \leq \delta \leq 30^\circ$ at $\Lambda_{\rm EW}$ may
result from $\delta = 0^\circ$ at $\Lambda^{}_{\rm FS}$
in the MSSM10 case. In the MSSM50 case even $|\delta| \simeq 90^\circ$
can be obtained at $\Lambda_{\rm EW}$, as shown in Table II.

\subsubsection{The running of the sum $\delta + \rho + \sigma$}

Eq. (11) leads us to the RGE of the sum of the three
CP-violating phases:
\begin{eqnarray}
\frac{{\rm d}}{{\rm d} t} (\delta + \rho + \sigma) & \approx &
\frac{y^{2}_{\tau}}{4 \pi^2} \frac{m^{2}_{1}}{\Delta m^{2}_{32}}
\frac{1}{s^{}_{13}} \left [ \frac{1}{2} c^{}_{12} s^{}_{12} c^{}_{23}
s^{}_{23}
\left ( s^{}_{(\delta + 2 \rho)} - s^{}_{(\delta + 2 \sigma)} \right )
\right. \nonumber\\[2mm]
& & ~~~  \left. - \left ( c^{}_{(\delta - \rho)} s^{}_{(\delta - \rho)}
s^{2}_{12} +c^{}_{(\delta - \sigma)} s^{}_{(\delta - \sigma)} c^{2}_{12}
\right ) \cos2\theta^{}_{23} s^{}_{13} \right. \nonumber\\[2mm]
& & ~~~ \left. + \left ( c^{}_{\rho} s^{}_{\rho} c^{2}_{12} +
c^{}_{\sigma} s^{}_{\sigma} s^{2}_{12} \right ) c^{2}_{23} s^{}_{13}
+ {\cal O} (s^{2}_{13}) \right ] \; .
\end{eqnarray}
Since the value of $\theta^{}_{13}$ is not small, the RGE running effect
on $(\delta + \rho + \sigma)$ is expected to be insignificant. In other words,
the sum of the three CP-violating phases approximately keeps unchanged
during the RGE evolution in the standard model or MSSM with a modest
$\tan\beta$. Our numerical analysis shows that $(\delta + \rho +
\sigma)$ changes less than $4^\circ$ when running from
$\Lambda^{}_{\rm FS}$ down to $\Lambda^{}_{\rm EW}$ in the MSSM10 case.
The stability of $(\delta + \rho + \sigma)$ against the radiative
corrections is quite impressive, unless $\tan\beta$ is sufficiently large.

\subsubsection{On the normal and inverted mass hierarchies}

In all the above discussions, we have assumed that the three neutrino
masses are nearly degenerate. For the purpose of completeness, here
we give a brief discussion on the RGE effects by considering
the hierarchical neutrino mass spectrum.
The radiative corrections to the three
mixing angles in both normal ($m^{}_1 < m^{}_2 < m^{}_3$) and
inverted ($m^{}_3 < m^{}_1 < m^{}_2$) neutrino mass hierarchies are less
significant than those in the case of a nearly degenerate neutrino mass
spectrum. Here we focus on two special but instructive
cases: i) the normal hierarchy with
$m^{}_{1} \simeq 0$ and ii) the inverted hierarchy with $m^{}_{3}
\simeq 0$. We choose $\theta^{}_{12} = 34^\circ$, $\theta^{}_{23} =
46^\circ$, $\theta^{}_{13} = 9^\circ$, $\delta = 90^\circ$, $\rho =
60^\circ$ and $\sigma = 30^\circ$ as the typical inputs at
$\Lambda^{}_{\rm EW}$ and study their RGE running behaviors to
$\Lambda^{}_{\rm FS}$. The corresponding outputs at
$\Lambda^{}_{\rm FS}$ in both the MSSM10 case and the MSSM50 case are
summarized in Table III. In the case of the normal hierarchy with
$m^{}_{1} \simeq 0$, the resulting values of the three mixing angles
at $\Lambda^{}_{\rm FS}$ are all close to their values at $\Lambda^{}_{\rm
EW}$, implying that the RGE running effects are insignificant.
As for the inverted hierarchy with $m^{}_{3} \simeq 0$,
the evolution of $\theta^{}_{23}$ and $\theta^{}_{13}$ is also
insignificant, but that of $\theta^{}_{12}$ is appreciable in the MSSM10
case and quite significant in the MSSM50 case.

Now let us make a brief summary. In order to obtain a
phenomenologically-favored neutrino mixing pattern at the
electroweak scale $\Lambda^{}_{\rm EW}$, we have examined
the corresponding mixing pattern at a superhigh-energy scale
$\Lambda^{}_{\rm FS}$ which might result from a certain flavor
symmetry. In the MSSM10 case the
values of $\theta^{}_{23}$ and $\theta^{}_{13}$ predicted at
$\Lambda^{}_{\rm FS}$ are always close to their running values
at $\Lambda^{}_{\rm EW}$, while the value of $\theta^{}_{12}$
at $\Lambda^{}_{\rm FS}$ can be somewhat smaller or larger than
its running value at $\Lambda^{}_{\rm EW}$.
In the MSSM50 case the allowed ranges of the three mixing angles at
$\Lambda^{}_{\rm FS}$ can be quite wide, as we have
discussed above. However, a crucial point is that a given
flavor symmetry model should be able to predict the appropriate
CP-violating phases at $\Lambda^{}_{\rm FS}$ in order to obtain the
appropriate mixing angles at $\Lambda^{}_{\rm EW}$ after the RGE
evolution. We find that it is in general impossible to generate
$\theta^{}_{13} \simeq 9^\circ$ at $\Lambda^{}_{\rm EW}$ from
$\theta^{}_{13} \simeq 0^\circ$ at $\Lambda^{}_{\rm FS}$
through the radiative corrections, unless some new degrees of freedom or
nontrivial running effects (such as the seesaw threshold effects
\cite{threshold effect}) are taken into account. This observation is
consistent with the discussions in Ref. \cite{Goswami2009}.

\section{An example: the Correlative Mixing Pattern}

In this section we consider the correlative neutrino mixing pattern
with $\theta^{}_{12} \simeq 35.3^\circ$, $\theta^{}_{23} = 45^\circ$ and
$\theta^{}_{13} \simeq 9.7^\circ$ \cite{correlative mixing pattern}.
The three mixing angles in this constant scenario satisfy two interesting
sum rules,
\begin{eqnarray}
&& \theta^{}_{12} + \theta^{}_{13} = \theta^{}_{23} \; , \nonumber \\
&& \theta^{}_{12} + \theta^{}_{13} +
\theta^{}_{23} = 90^\circ \; .
\end{eqnarray}
The latter sum rule is geometrically illustrated
in FIG. 3. The corresponding lepton flavor mixing matrix is
\cite{correlative mixing pattern}
\begin{equation}
V = \left( \matrix{ \displaystyle \frac{\sqrt{2} + 1}{3} &
\displaystyle \frac{\sqrt{2} + 1}{3\sqrt{2}} &  \displaystyle
\frac{\sqrt{2} - 1}{\sqrt{6}} e^{-i\delta} \cr\cr - \displaystyle
\frac{1}{\sqrt{6}} - \displaystyle \frac{\sqrt{2} - 1}{3\sqrt{2}}
e^{i\delta} & ~ \displaystyle \frac{1}{\sqrt{3}} - \displaystyle
\frac{\sqrt{2} - 1}{6} e^{i\delta} & \displaystyle \frac{\sqrt{2}
+ 1}{2\sqrt{3}} \cr\cr ~ \; \displaystyle \frac{1}{\sqrt{6}} -
\displaystyle \frac{\sqrt{2} - 1}{3\sqrt{2}} e^{i\delta} & -
\displaystyle \frac{1}{\sqrt{3}} - \displaystyle \frac{\sqrt{2}
- 1}{6} e^{i\delta} & \displaystyle \frac{\sqrt{2} + 1}{2\sqrt{3}} \cr
} \right)
\left ( \matrix{ e^{i \rho} & 0 & 0 \cr 0 & e^{i \sigma} & 0
\cr 0 & 0 & 1 \cr } \right ) \; ,
\end{equation}
which might be derived from a certain underlying flavor
symmetry in a neutrino mass model at a superhigh-energy scale.

Compared with the best-fit values of the three mixing angles at
$\Lambda^{}_{\rm EW}$ given in Ref. \cite{Global Fit}, the three mixing
angles in this correlative mixing pattern at $\Lambda^{}_{\rm FS}$
have to receive comparably small radiative corrections during
their RGE evolution. As we have mentioned in the last
section, this requirement can easily be achieved in the MSSM10 case
provided the condition $(\rho -\sigma) \simeq \pm 90^\circ$ is satisfied
for a nearly degenerate neutrino mass spectrum. Such a condition
is unnecessary if the neutrino mass spectrum has a strong
hierarchy.

To illustrate, we study the RGE evolution of $V$ in Eq. (16) and
present our results in Table IV. We input $\delta = -68^\circ$,
$\rho = 13^\circ$ and $\sigma = 115^\circ$ at $\Lambda_{\rm FS}$ for
example, and then obtain the phenomenologically-favored results
$\theta^{}_{12} = 34.52^\circ$,
$\theta^{}_{23} = 45.98^\circ$ and $\theta^{}_{13} = 8.83^\circ$ at
$\Lambda_{\rm EW}$ after the radiative corrections. If the three
CP-violating phases can be specified in a given flavor symmetry
model at $\Lambda^{}_{\rm FS}$, however, the running behaviors of
the three mixing angles will be more restrictive.

Given the RGEs in the framework of the standard model \cite{Majorana
RGE,RGE}, the evolution
of $\theta^{}_{12}$, $\theta^{}_{13}$ and $\theta^{}_{23}$ in the
correlative mixing pattern is insignificant but their running directions
may be opposite to those in the MSSM case (depending on the CP-violating
phases). If the experimental error bars of the three mixing angles
turn to be sufficiently small in the future, it might be possible
to see which framework is more suitable for the radiative corrections
to the correlative neutrino mixing pattern.

\section{Summary}

In view of $\theta^{}_{13} \simeq 9^\circ$ as observed in the recent
Daya Bay and RENO experiments, we have reexamined the
radiative corrections to the lepton flavor mixing matrix for both
Dirac and Majorana neutrinos by considering both nearly
degenerate and strongly hierarchical neutrino mass spectra in the
framework of the MSSM. Two typical values of $\tan\beta$ (i.e.,
10 and 50) have been taken in our numerical calculations. We conclude
that it is difficult to generate $\theta^{}_{13} \simeq 9^\circ$
at $\Lambda_{\rm EW}$ from $\theta^{}_{13} \simeq 0^\circ$ at
$\Lambda_{\rm FS}$ through the radiative corrections, unless a
sufficiently large value of $\tan\beta$ is assumed or the seesaw
threshold effects or some new degrees of freedom are taken into
account. Therefore, we argue that it is more natural for a
flavor symmetry model to predict a relatively large $\theta^{}_{13}$
at $\Lambda_{\rm FS}$. To illustrate this point, we have briefly discussed
the correlative mixing pattern with $\theta^{}_{12} \simeq 35.3^\circ$, $\theta^{}_{23} = 45^\circ$ and $\theta^{}_{13} \simeq 9.7^\circ$ as
an example of this kind.

Let us remark that fixing $\theta^{}_{13} \simeq 9^\circ$ and taking
the correlative neutrino mixing pattern in our RGE analysis just serve for
illustration. It is certainly possible to generate the values of
$\theta^{}_{13}$ within the present experimental limits (but not
necessarily close to $9^\circ$) from $\theta^{}_{13} \simeq 0^\circ$
at a superhigh-energy scale in the MSSM with an appropriate value
of $\tan\beta$ or with the help of some new degrees of freedom. However,
we stress that $\theta^{}_{13}$ might initially be nonzero and its
appreciable value might have a significant impact on the running
behaviors of the other two mixing angles and CP-violating phases.

Our study clearly shows that a measurement of the Dirac CP-violating phase
$\delta$ in the forthcoming long-baseline oscillation experiments
and any experimental information about the Majorana CP-violating
phases $\rho$ and $\sigma$ are extremely important,
so as to distinguish one flavor symmetry model from another through
their different sensitivities to the radiative corrections. This
observation makes sense in particular after the experimental
errors associated with the neutrino mixing parameters are
comparable with or smaller than the magnitudes of their respective
RGE running effects.

\vspace{0.3cm}


We are grateful to P. Minkowski, W. Rodejohann and H. Zhang for
useful discussions.
The work of S.L. is supported in part by the National Basic Research
Program (973 Program) of China under Grant No. 2009CB824800, the
National Natural Science Foundation of China under Grant No.
11105113, the Fujian Provincial Natural Science Foundation under
Grant No. 2011J05012 and the China Postdoctoral Science Foundation
funded project under Grant No. 201104340. The work of Z.Z.X.
is supported in part by the National Natural Science
Foundation of China under grant No. 11135009.


\newpage

\begin{figure}
\centering
\includegraphics[bb = 230 149 380 824,scale=0.8]{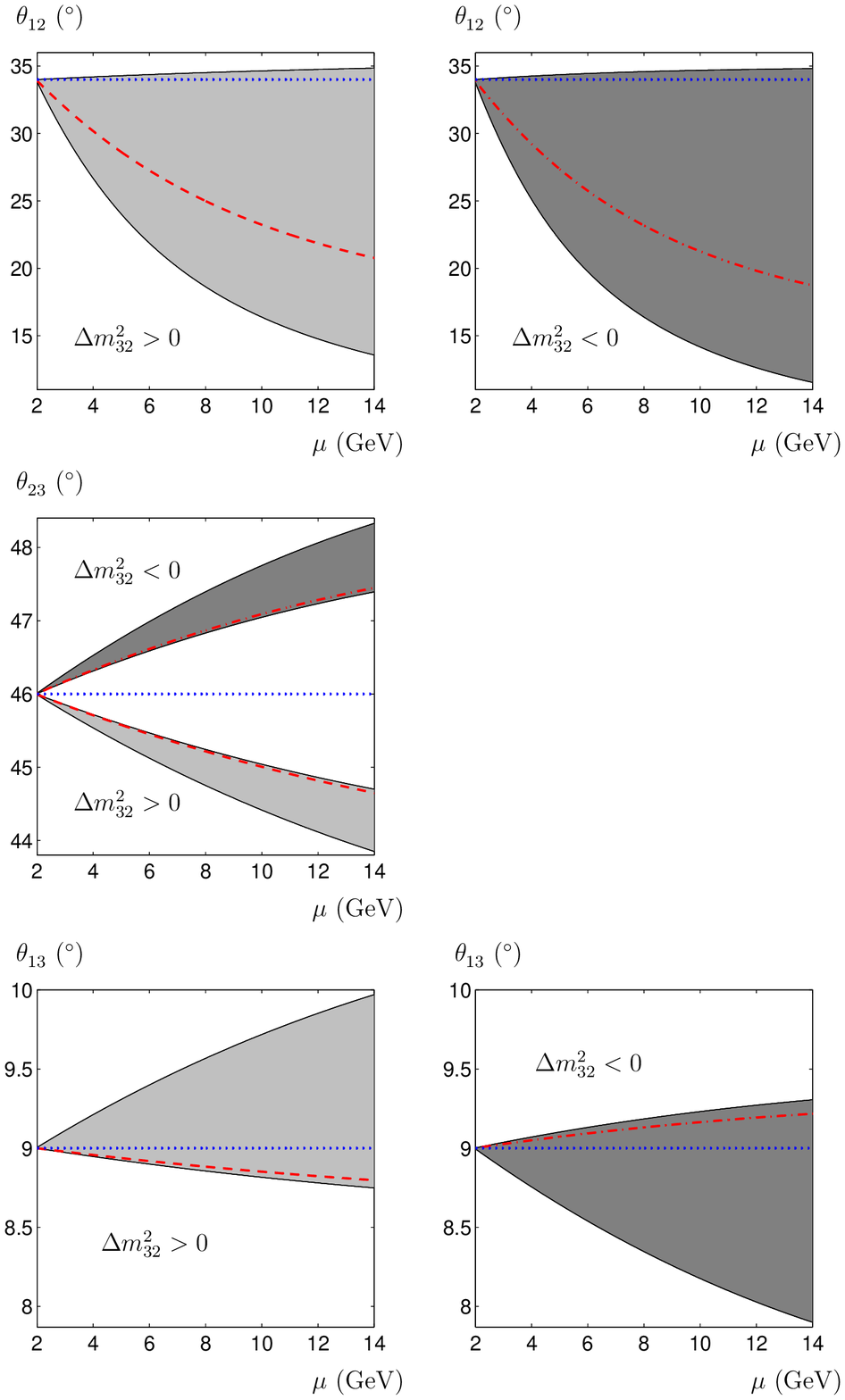}
\caption{Running behaviors of the three mixing angles
$\theta^{}_{12}$, $\theta^{}_{23}$ and $\theta^{}_{13}$ from
$\Lambda^{}_{\rm FS} \sim 10^{14}$ GeV to $\Lambda^{}_{\rm EW} \sim
10^{2}$ GeV in the MSSM10 case. We have taken
$\theta^{}_{12} = 34^\circ$, $\theta^{}_{23} = 46^\circ$,
$\theta^{}_{13} = 9^\circ$, $\delta = 90^\circ$ and $\sigma =
30^\circ$ as the typical inputs at $\Lambda^{}_{\rm EW}$ for
the Majorana neutrinos, and allowed $\rho$ to vary from
$0^\circ$ to $180^\circ$.  The gray areas show the possible ranges
of the three mixing angles at $\mu > \Lambda^{}_{\rm EW}$. The
dashed ($\Delta m^{2}_{32} > 0$) and dotted-dashed ($\Delta
m^{2}_{32} < 0$) lines represent the corresponding running behaviors
of the Dirac neutrinos with the same inputs at $\Lambda^{}_{\rm EW}$.}
\end{figure}

\begin{figure}
\centering
\includegraphics[bb = 230 140 380 840,scale=0.8]{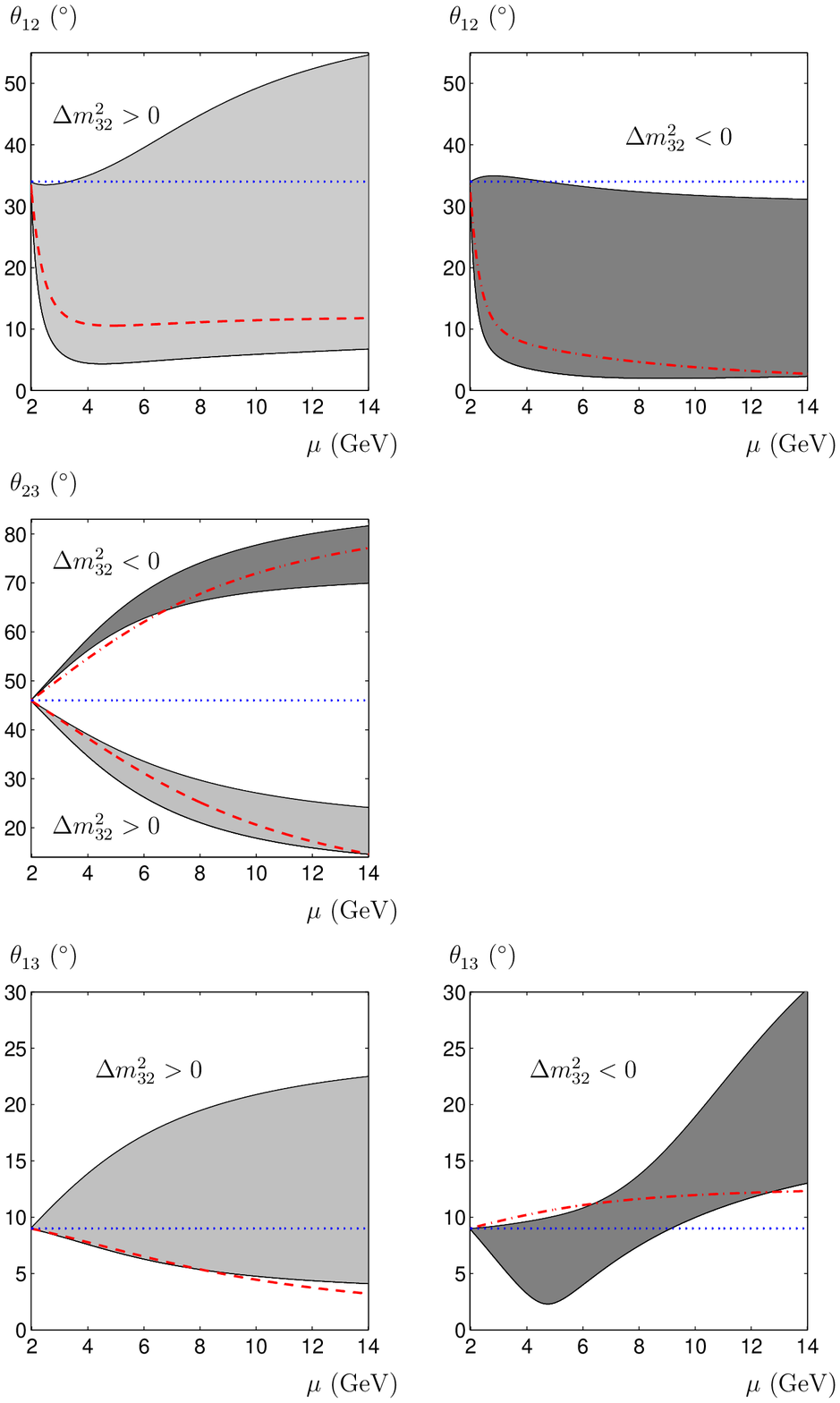}
\caption{Running behaviors of the three mixing angles
$\theta^{}_{12}$, $\theta^{}_{23}$ and $\theta^{}_{13}$ from
$\Lambda^{}_{\rm FS} \sim 10^{14}$ GeV to $\Lambda^{}_{\rm EW} \sim
10^{2}$ GeV in the MSSM50 case. We have taken
$\theta^{}_{12} = 34^\circ$, $\theta^{}_{23} = 46^\circ$,
$\theta^{}_{13} = 9^\circ$, $\delta = 90^\circ$ and $\sigma =
30^\circ$ as the typical inputs at $\Lambda^{}_{\rm EW}$ for
the Majorana neutrinos, and allowed $\rho$ to vary from
$0^\circ$ to $180^\circ$.  The gray areas show the possible ranges
of the three mixing angles at $\mu > \Lambda^{}_{\rm EW}$. The
dashed ($\Delta m^{2}_{32} > 0$) and dotted-dashed ($\Delta
m^{2}_{32} < 0$) lines represent the corresponding running behaviors
of the Dirac neutrinos with the same inputs at $\Lambda^{}_{\rm EW}$.}
\end{figure}

\begin{figure}
\centering
\includegraphics[bb = 230 460 380 760,scale=0.8]{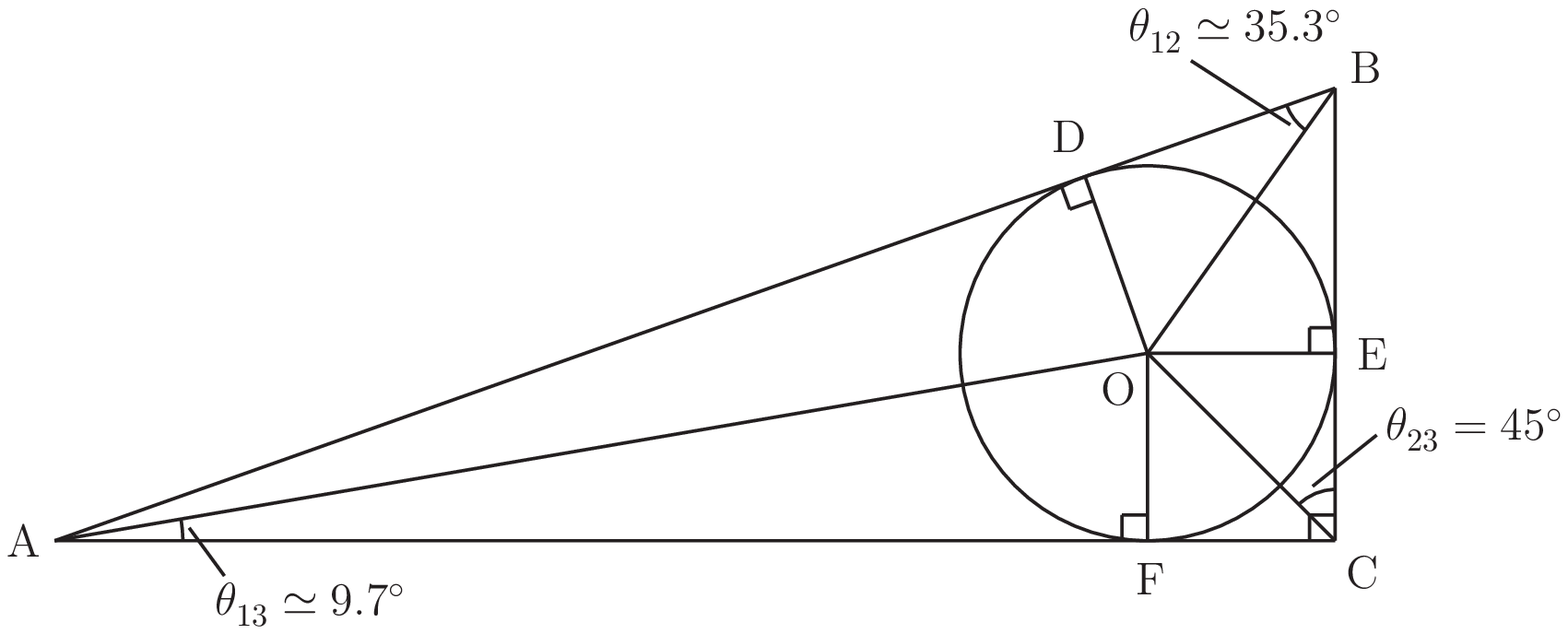}
\caption{A geometrical description of the sum rules
$\theta^{}_{12} + \theta^{}_{13} = \theta^{}_{23}$ and
$\theta^{}_{12} + \theta^{}_{13} + \theta^{}_{23} = 90^\circ$ for
the correlative neutrino mixing pattern in terms of the inner angles
of the right triangle
$\triangle\rm ABC$.}
\end{figure}

\begin{table}
\caption{Examples of the radiative generation of the CP-violating phases
from $\Lambda^{}_{\rm FS} \sim 10^{14}$ GeV to $\Lambda^{}_{\rm EW} \sim
10^2$ GeV in the MSSM10 case.}
\begin{center}
\begin{tabular}{llllllll}
~ Parameter & ~~~~~~ & \multicolumn{2}{l}{Example I} & ~~~ &
\multicolumn{2}{l}{Example II} \\
\cline{3-4}\cline{6-7}
& & Input ($\Lambda^{}_{\rm FS}$) ~~~ & Output ($\Lambda^{}_{\rm EW}$) ~~~
& & Input ($\Lambda^{}_{\rm FS}$) ~~~ & Output ($\Lambda^{}_{\rm EW}$) ~~~ \\
\hline
~ $m^{}_{1} ~ ({\rm eV} )$ && 0.227 & 0.200 && 0.227 & 0.200\\
~ $\Delta m^{2}_{21} ~ ( 10^{-5} ~{\rm eV}^2 )$ && 20.47 & 7.59 && 21.09 & 7.59 \\
~ $\Delta m^{2}_{31} ~ ( 10^{-3} ~{\rm eV}^2 )$ && 3.22 & 2.40 && 3.22 & 2.40 \\
~ $\theta^{}_{12}$ && $25.1^\circ$ & $33.97^\circ$ && $23.7^\circ$ & $33.95^\circ$ \\
~ $\theta^{}_{23}$ && $43.5^\circ$ & $45.99^\circ$ && $44.7^\circ$ & $45.97^\circ$ \\
~ $\theta^{}_{13}$ && $7.9^\circ$ & $8.78^\circ$ && $9.0^\circ$ & $8.85^\circ$ \\
~ $\delta$ && $0^\circ$ & $26.05^\circ$ && $0^\circ$ & $-26.50^\circ$ \\
~ $\rho$ && $55^\circ$ & $36.52^\circ$ && $0^\circ$ & $17.92^\circ$ \\
~ $\sigma$ && $0^\circ$ & $-5.65^\circ$ && $50^\circ$ & $55.44^\circ$ \\
~ $\delta + \rho + \sigma$ && $55^\circ$ & $56.92^\circ$ && $50^\circ$
& $46.86^\circ$ \\
\end{tabular}
\end{center}
\end{table}

\begin{table}
\caption{Examples of the radiative generation of the CP-violating phases
from $\Lambda^{}_{\rm FS} \sim 10^{14}$ GeV to $\Lambda^{}_{\rm EW} \sim
10^2$ GeV in the MSSM50 case.}
\begin{center}
\begin{tabular}{llllllll}
~ Parameter & ~~~~~~ & \multicolumn{2}{l}{Example I} & ~~~ &
\multicolumn{2}{l}{Example II} \\
\cline{3-4}\cline{6-7}
& & Input ($\Lambda^{}_{\rm FS}$) ~~~ & Output ($\Lambda^{}_{\rm EW}$) ~~~
& & Input ($\Lambda^{}_{\rm FS}$) ~~~ & Output ($\Lambda^{}_{\rm EW}$) ~~~ \\
\hline
~ $m^{}_{1} ~ ({\rm eV} )$ && 0.245 & 0.200 && 0.245 & 0.200\\
~ $\Delta m^{2}_{21} ~ ( 10^{-5} ~{\rm eV}^2 )$ && 160.20 & 7.57 && 671.70 & 7.61 \\
~ $\Delta m^{2}_{31} ~ ( 10^{-3} ~{\rm eV}^2 )$ && 16.03 & 2.40 && 10.10 & 2.39 \\
~ $\theta^{}_{12}$ && $10.86^\circ$ & $34.01^\circ$ && $15.77^\circ$ & $34.03^\circ$ \\
~ $\theta^{}_{23}$ && $7.50^\circ$ & $46.06^\circ$ && $44.03^\circ$ & $46.00^\circ$ \\
~ $\theta^{}_{13}$ && $2.88^\circ$ & $9.04^\circ$ && $12.99^\circ$ & $9.01^\circ$ \\
~ $\delta$ && $0^\circ$ & $91.48^\circ$ && $0^\circ$ & $87.93^\circ$ \\
~ $\rho$ && $100^\circ$ & $9.83^\circ$ && $0^\circ$ & $-62.65^\circ$ \\
~ $\sigma$ && $0^\circ$ & $-4.75^\circ$ && $95^\circ$ & $84.09^\circ$ \\
\end{tabular}
\end{center}
\end{table}

\begin{table}
\caption{Radiative corrections to the neutrino masses and flavor
mixing parameters from $\Lambda^{}_{\rm EW} \sim 10^{2}$ GeV to
$\Lambda^{}_{\rm FS} \sim 10^{14}$ GeV in the MSSM with $\tan\beta =10$
or $50$.}
\begin{center}
\begin{tabular}{lllllllll}
~ Parameter & \multicolumn{3}{l}{Normal hierarchy with $m^{}_{1} \simeq 0$} &  &
\multicolumn{3}{l}{Inverted hierarchy with $m^{}_{3} \simeq 0$} \\
\cline{2-4}\cline{6-8}
& Input ($\Lambda^{}_{\rm EW}$) & \multicolumn{2}{l}{Output
($\Lambda^{}_{\rm FS}$)} & & Input ($\Lambda^{}_{\rm EW}$)
& \multicolumn{2}{l}{Output ($\Lambda^{}_{\rm FS}$)} \\
\cline{3-4}\cline{7-8}
& & MSSM10 ~ & MSSM50
& & & MSSM10 ~ & MSSM50 \\
\hline
~ $m^{}_{1} ({\rm eV} )$ & ~ $10^{-6}_{} $ & $1.13 \times 10^{-6}_{}$ & $1.24 \times 10^{-6}_{}$ && ~ 0.049 & 0.053 & 0.059 \\
~ $\Delta m^{2}_{21} ( 10^{-5} ~{\rm eV}^2 )$ & ~ 7.59 & 9.80 & 12.23 && ~ 7.59 & 9.28 & 55.10 \\
~ $\Delta m^{2}_{31} ( 10^{-3} ~{\rm eV}^2 )$ & ~ 2.40 & 3.10 & 3.97 && ~ -2.40 & -2.82 & -3.50 \\
~ $\theta^{}_{12}$ & ~ $34.0^\circ$ & $33.98^\circ$ & $33.28^\circ$ && ~ $34.0^\circ$ & $31.76^\circ$ & $14.06^\circ$ \\
~ $\theta^{}_{23}$ & ~ $46.0^\circ$ & $45.96^\circ$ & $44.32^\circ$ && ~ $46.0 ^\circ$ & $46.04^\circ$ & $47.75^\circ$ \\
~ $\theta^{}_{13}$ & ~ $9.0^\circ$ & $9.00 ^\circ$ & $8.99^\circ$ && ~ $9.0^\circ$ & $9.01^\circ$ & $9.26^\circ$ \\
~ $\delta$ &~ $90^\circ$ & $90.01^\circ$ & $90.19^\circ$ && ~ $90^\circ$ & $87.10^\circ$ & $29.18^\circ$ \\
~ $\rho$ & ~ $60^\circ$ & $60.02^\circ$ & $60.97^\circ$ && ~ $60^\circ$ & $62.04^\circ$ & $113.93^\circ$ \\
~ $\sigma$ & ~ $30^\circ$ & $30.00^\circ$ & $29.95^\circ$ && ~ $30^\circ$ & $30.86^\circ$ & $39.05^\circ$ \\
\end{tabular}
\end{center}
\end{table}

\begin{table}
\caption{Radiative corrections to the correlative neutrino mixing pattern
from $\Lambda^{}_{\rm FS} \sim 10^{14}$ GeV to $\Lambda^{}_{\rm EW} \sim
10^2$ GeV in the MSSM10 case.}
\begin{center}
\begin{tabular}{llllllll}
~ Parameter & ~~~~~~ & Input ($\Lambda^{}_{\rm FS}$) ~~~ & Output
($\Lambda^{}_{\rm EW}$) \\
\hline
~ $m^{}_{1} ~ ({\rm eV} )$ && 0.227 & 0.200 \\
~ $\Delta m^{2}_{21} ~ ( 10^{-5} ~{\rm eV}^2 )$ && 15.72 & 7.59 \\
~ $\Delta m^{2}_{31} ~ ( 10^{-3} ~{\rm eV}^2 )$ && 3.19 & 2.40 \\
~ $\theta^{}_{12}$ && $35.3^\circ$ & $34.52^\circ$ \\
~ $\theta^{}_{23}$ && $45^\circ$ & $45.98^\circ$ \\
~ $\theta^{}_{13}$ && $9.7^\circ$ & $8.83^\circ$ \\
~ $\delta$ && $-68^\circ$ & $-80.88^\circ$ \\
~ $\rho$ && $13^\circ$ & $19.64^\circ$ \\
~ $\sigma$ && $115^\circ$ & $118.03^\circ$ \\
~ $\delta + \rho + \sigma$ && $60^\circ$ & $56.79^\circ$ \\
\end{tabular}
\end{center}
\end{table}

\end{document}